\documentstyle[12pt]{article}
\newcommand{\newsection}[1]{
\addtocounter{section}{1}
\setcounter{equation}{0}
\setcounter{subsection}{0}
\addcontentsline{toc}{section}{\protect
\numberline{\arabic{section}}{{\rm #1}}}
\vglue .6cm
\pagebreak[3]
\noindent{\bf  \thesection. #1}\nopagebreak[4]\par\vskip .3cm}
\newcommand{\newsubsection}[1]{
\addtocounter{subsection}{1}
\addcontentsline{toc}{subsection}{\protect
\numberline{\arabic{section}.\arabic{subsection}}{#1}}
\vglue .4cm
\pagebreak[3]
\noindent{\it \thesubsection. #1}\nopagebreak[4]\par\vskip .3cm}

\renewcommand{\theequation}{\thesection.\arabic{equation}}

\newcommand{\ben}{\begin{enumerate}}
\newcommand{\een}{\end{enumerate}}
\newlength{\extraspace}
\newlength{\extraspaces}
\newcounter{dummy}
\newcommand{\bc}{\begin{center}}
\newcommand{\ec}{\end{center}}
\newcommand{\be}{\begin{equation}
\addtolength{\abovedisplayskip}{\extraspaces}
\addtolength{\belowdisplayskip}{\extraspaces}
\addtolength{\abovedisplayshortskip}{\extraspace}
\addtolength{\belowdisplayshortskip}{\extraspace}}
\newcommand{\ee}{\end{equation}}

\newcommand{\ba}{\begin{eqnarray}
\addtolength{\abovedisplayskip}{\extraspaces}
\addtolength{\belowdisplayskip}{\extraspaces}
\addtolength{\abovedisplayshortskip}{\extraspace}
\addtolength{\belowdisplayshortskip}{\extraspace}}
\newcommand{\ea}{\end{eqnarray}}

\newcommand{\ban}{\begin{eqnarray*}
\addtolength{\abovedisplayskip}{\extraspaces}
\addtolength{\belowdisplayskip}{\extraspaces}
\addtolength{\abovedisplayshortskip}{\extraspace}
\addtolength{\belowdisplayshortskip}{\extraspace}}
\newcommand{\ean}{\end{eqnarray*}}
\newcommand{\baa}{                         
\addtocounter{equation}{1}
\setcounter{dummy}{\value{equation}}
\setcounter{equation}{0}
\renewcommand{\theequation}{\thesection.\arabic{dummy}\alph{equation}}
\begin{eqnarray}
\addtolength{\abovedisplayskip}{\extraspaces}
\addtolength{\belowdisplayskip}{\extraspaces}
\addtolength{\abovedisplayshortskip}{\extraspace}
\addtolength{\belowdisplayshortskip}{\extraspace}}
\newcommand{\eaa}{                                       
\end{eqnarray}
\setcounter{equation}{\value{dummy}}
\renewcommand{\theequation}{\thesection.\arabic{equation}}}

\input epsf

\newcounter{fignum}
\newcounter{tabel}

\newcounter{tabnum}
\newcounter{xxx}
\newcommand{\bl}{\begin{list}{({\it\roman{xxx}})}{\usecounter{xxx}}}
\newcommand{\el}{\end{list}}
\newcommand{\ppt}[1]{{\partial \over \partial t}}            
\newcommand{\ppx}[1]{{\partial \over \partial x}}            
\newcommand{\pqt}[1]{{\partial^2 \over \partial t^2}}            
\newcommand{\pqx}[1]{{\partial^2  \over \partial x^2}}            
\newcommand{\twomatrix}[4]{{\left(\begin{array}{cc}#1 & #2\\
#3 & #4 \end{array}\right)}}

\renewcommand{\l}{\langle}

\newcommand{\th}{^{\mit th}}

\newcommand{\del}{\partial}
\def\a{\alpha} 
\def\b{\beta} 
\def\g{\gamma} 

\def\e{\epsilon}

\def\th{\theta}

\def\l{\lambda}

\def\s{\sigma}

\def\<{\langle}
\def\>{\rangle}

\newfont{\gothic}{eufm10 scaled\magstep1}

\newcommand{\pslash}{{p \hspace{-5pt} \slash}}

\renewcommand{\hat}{\widehat}

\newcounter{problems}

\begin{document}
\begin{titlepage}
\begin{flushleft}
\today
\end{flushleft}
\begin{center}
{\LARGE \bf Left-Ideals, Dirac fermions and $SU(2)$-Flavour}
\end{center}
\vskip 2cm
\begin{center}
\mbox{F.M.C.\ Witte}\\
{
\it
Julius School of Physics and Astronomy, 
Utrecht University\\
Leuvenlaan 4, 3584 CE, Utrecht\\
Netherlands}
\end{center}
\vskip 1.5cm
\begin{abstract}
In this paper I reconsider the use of the left ideals of the even-grade subalgebra of spacetime algebra to describe fermionic excitations. When interpreted as rotors the general elements of an even-grade left-ideal describe massless particles in chiral flavour doublets. To study the application of these ideas to the standard Dirac formalism I construct a $2 \times 2$-matrix representation with bivector insertions for the Dirac algebra. This algebra has four ideals, and this approach clarifies how the identification of Dirac $\g_{\mu}$-matrices with orthonormal basisvectors ${\bf e}_{\nu}$ annihilates half of the ideals. For one possible choice of this mapping the remaining ideals the chiral left- and righthanded components of the fermion coincide with the even- and odd elements of spacetime algebra.
\end{abstract}
\end{titlepage}
\newpage

\newsection{Introduction}
Dirac's succesful attempt at making one-particle quantum mechanics compatible with special relativity drove him to the introduction of a set of matrices, ${\hat{\g}}^{\mu}$, whose algebra identifies them as generators of a Clifford algebra. Their appearance in the Dirac equation closely tied them to the notion of vector-components. If Dirac's gamma-matrices were taken to transform under a spinorial representation $\Lambda$ of a lorentz transformation $L$ as
\be
({\hat{\g}}^{\mu})' = \Lambda {\hat{\g}^{\mu}} \Lambda^{\dagger} \ , 
\ee
then contractions of these matrices with components of a spacetime vector like, for example, energy-momentum ${\bf p}$ could be interpreted as Lorentz scalars
\be
\pslash = \sum_{\mu = 0}^{3} {\hat{\g}}^{\mu} p_{\mu} \ .
\ee
This invariance of $\pslash$ can also be intepreted as the invariance of a vector under {\it passive} Lorentztransformations. The Dirac {\it spinors} are 4d complex dimensional collumn matrices and they could infact be regarded as an arbitrary collum in a $4 \times 4$ matrix with zero entries in the remaining three collumns. This construction is possible due to the fact that Lorentz transformations act {\it one-sided} on Dirac spinors. Left-multiplication by Dirac-matrices preserves the linear subspace represented by a single collumn in a matrix. The Dirac- matrix representation of spinors allows for {\it four} {\it left-ideals} covering 32 {\it real} degrees of freedom in total.

When considering {\it active} lorentz transformations the $p_{\mu}$ retain their interpretation as components of the momentum-vector but the gamma-matrices can be taken to be a representation of orthonormal reciprocal basis vectors ${\bf e}^{\mu}$. Both $\g^{\mu}$ as well as ${\bf e}^{\mu}$, in the geometric algebra of spacetime (STA), satisfy the defining relations of a Clifford algebra. This suggests that there exists a linear map
\be
\g^{\mu} \rightarrow {\bf e}^{\mu} \ .
\ee
Hestenes constructed the notion of geometric algebras, in particular their spacetime version STA, and the corresponding reformulation of the Dirac equation \cite{hest}. The use of geometric algebra and left-ideals in the construction of Dirac spinors was also discussed in \cite{spinors}. Remarkable points in that reformulation are the association of even-grade multi-vectors, i.e. {\it weighed rotors}, to Dirac wavefunctions and the identification of the unit-imaginary with the spin-plane bivector \cite{sta}. A multitude of  applications of spacetime algebra have since been discussed in the literature, including studies on the interpretation of the Dirac equation as well as applications to classical electromagnetism, and found their way into recent textbooks \cite{staphys}. Its application to the study of gravity and of fermion fields in gravitational backgrounds has only begun recently \cite{gagrav}, as has the appplication of these ideas to the analysis of the relativistic dynamics of extended objects \cite{extobj}. 

The idea behind such a line of research is, to my mind, primarilly to uncover the aspects of {\it spacetime} geometry embedded in elementary particle physics, going beyond the thought that the mere presence of fibre bundles is enough of identifiable geometry. Translating the Standard Model into spacetime algebra should reveal how to "geometrize" the quantum theory of the Standard Model, in contrast to the more standard endeavour to "quantize" the geometry of spacetime.

Despite a fair share of interpretative, and technical, success of this programme, in the spinorial wavefunctions of the electron the so-called $\b$-parameter has eluded definite geometric-physical interpretation \cite{relqm} so far. Although it has been argued that there must be some connection between STA and the structure of electroweak interactions \cite{sm} the exact relationships have not yet been revealed and alternative approaches based on geometric algebras in higher-dimensions have been put forward \cite{smplus}. Sometimes regarded as a postdiction, the fact that the (real) even-grade subalgebra of spacetime algebra only has two left-ideals whereas the Dirac algebra has four could be cause for concern as it does not seem to result from any specific {\it physical} constraint on the algebra. Furthermore the left- and righthanded chiral components of a Dirac fermion enter into the electroweak interactions in very distinct ways making them appear as very different degrees of freedom rather than as different components of a single particle. It seems odd, in a way, that a Dirac spinor describing both the left-handed and the right-handed electron should somehow translate into a {\it single-particle} rotor instead of a pair of two of these rotors.

In this paper I will provide arguments for the claim that the Dirac equation should be taken to describe the dynamics of {\it two} excitations, i.e. encompass a {\it bi-rotor}. This result follows from a re-analysis of the algebraic structure of left-ideals in STA and their use as wavefunctions. As it turns out, the typical left-ideal in the even-grade subalgebra of STA (called a Pauli-ideal for short) describes {\it massless} excitations with spin. When we identify the Dirac $\g$-matrices as $2 \times 2$ matrices with timelike bivectors as entries a Dirac equation emerges that contains four weighed rotors, combined in two parity-pairs, each of the four rotors representing a Dirac-matrix ideal with 8 real degrees of freedom. Every Dirac matrix ideal consists of two Pauli ideals with four real degrees of freedom each. Each of these two Pauli ideals is naturally associated with a component of an $SU(2)$-flavour doublet. The action of the flavour $SU(2)$ symmetry on these rotors is straightforward and can be understood as resulting from a symmetry under particle-frame rotations of the particle. As a result also the Higgs field can be given a spacetime geometric interpretation.

This paper is organised as follows. In section 2 we give a brief outline of the ideas presented here within the framework of 3d geometric algebra. The purpose here is threefold; firstly to present some of the basic notions neccesary to follow the main line of the paper, secondly to show that key ingredients for this work are already present in 3d geometric algebra and thirdly to make explicit which structural failures are remedied by the transition to spacetime. In section 3 I discus the Pauli ideals in STA and their intepretations as wavefunctions for massless excitations. Finally in section 4 I construct the Dirac spinors, using the chiral representation for definiteness, and discuss their classical and quantum mechanical interpretation. Section 5 closes this paper with a summary and a brief discussion of the relevance of these results for the study of the geometric nature of electroweak interactions and the role of spin in a gravitational context.

\newsection{Pauli Ideals of 3d Geometric Algebra}
Given a vectorspace we define a geometric product as an associative and distributive product such that the square of a vector becomes a scalar. As a result the geometric product ${\bf a b}$ between to vectors ${\bf a}$ and ${\bf b}$ falls apart into a scalar and bivector according to
\ba
{\bf a} \cdot {\bf b} & = & \frac{1}{2}({\bf a b} + {\bf b a}) \ , \nonumber \\
{\bf a} \wedge {\bf b} & = & \frac{1}{2}({\bf a b} - {\bf b a}) \ , \\
{\bf a} {\bf b} & = & {\bf a} \cdot {\bf b}  + {\bf a} \wedge {\bf b} \ . \nonumber
\ea
Orthonormal basis vectors ${\bf e}_{j}$ anti-commute under the geometric product and thus satisfy
\be
{\bf e}_{i} {\bf e}_{j} + {\bf e}_{i} {\bf e}_{j} = 2 \eta_{ij} \ ,
\ee
where $\eta_{ij}$ is the metric of the vectorspace involved. This last equation is typically the defining equation for a Clifford algebra and matrix algebras can usually be found that satisfy it. Although this assures the existence of geometric algebras, there is no need to assign any specific physical importance to the matrix representation itself. One possible matrix-representation for the geometric algebra in three dimensional euclidean space, $GA_{3}$, is the Pauli algebra. Let the unit-pseudoscalar of a geometric algebra be denoted by $i$. Then in 3d euclidean space we have
\be
i = {\bf e}_{1}{\bf e}_{2}{\bf e}_{3} \ ,
\ee
and this pseudoscalar commutes with all vectors. It is then easy to check that these basis-vectors indeed satisfy the same commutation relations as the Pauli matrices.
\be
[ {\bf e}_{i} , {\bf e}_{j} ] = - i \e_{ijk} {\bf e}_{k} \ .
\ee
The matrix algebra of Pauli matrices is wellknown for its applications in quantum physics, where it is usually associated with the spin-operators for a spin $\frac{\hbar}{2}$ particle. But here we see it can also be used to represent basis-vectors in 3d.

Rotations can be represented by the action of even-grade multivectors $R$ satisfying $R \tilde{R} = 1$, where ${\tilde{R}}$ denotes the same multivector but with all products of vectors in reversed order. In general a multivector will transform under rotations as
\be
M' = R M {\tilde{R}} \ .
\ee
Rotors are generated by bivectors. For example, a rotation about an angle $\a$ in the plane ${\bf e}_{2}{\bf e}_{3}$ is described by the rotor 
\be
R_{23}(\a) = \exp{(-\frac{\a}{2} {\bf e}_{2} {\bf e}_{3})} \ .
\ee
Note that for $\a = 2 \pi$ we find $R_{23} = -1$! Rotors show spinorial behaviour which is because they transform according to a {\it one-sided } rule. As a generalisation every multivector $N$ transforming under rotations represented by the rotor $R$ as 
\be
N'= R N \ ,
\ee
is {\it spinorial} in nature. 

\newsubsection{Pauli ideals}
A left-ideal of $GA_{3}$ is a subset $I$ of multivectors such that
\be
A \in GA_{3} \ , B \in I \rightarrow A B \in I \ ,
\ee
is true. To find such ideals it is sufficient that they have this property with respect to the multiplication from the left by the generating vectors of some basis. A good starting point is an {\it eigen-multivector} of one of the basis vectors. Let us choose to start with ${\bf e}_{3}$, an obvious choice for the eigen multivectors is
\be
E_{\pm 3} = \frac{1}{2} ( 1 \pm {\bf e}_{3} ) \ .
\ee
That satisfies
\be
{\bf e}_{3} E_{\pm 3} = E_{\pm 3} {\bf e}_{3} = \pm E_{\pm 3} \ .
\ee
$E_{+3}$ and $E_{-3}$ generate different ideals. Using the unit-pseudoscalar we have
\ba
{\bf e}_{1} E_{\pm 3} & = & \frac{1}{2} ( {\bf e}_{1} \mp i {\bf e}_{2} ) \equiv D_{\pm 3} \ , \nonumber \\
{\bf e}_{2} E_{\pm 3} & = & i D_{\pm 3} \ , \\
{\bf e}_{1} D_{\pm 3} & = & -i E_{\pm 3} \ , \nonumber \\
{\bf e}_{3} D_{\pm 3} & = & \mp D_{\pm 3} \ . \nonumber 
\ea
The ideals have two generators, but as all commute with $i$ we may expand any element of the ideal in terms of coefficients which have scalar and pseudoscalar parts, i.e. behave as complex numbers. We can write
\be
\Psi_{\pm} = a_{\pm} E_{\pm 3} + b_{\pm} D_{\pm 3} \ ,
\ee
where the coefficients are scalar-pseudoscalar combinations like $a_{+} = a_{1} + i a_{2}$. 

Now consider applying this procedure to spinorial objects. The decomposition in ideals is trivially invariant under rotations. This raises the question whether they are in some way related to a decomposition of fermionic wavefunctions. If we compute bilinear forms, typical ingredients of quantum mechanical expectationvalues, we find, for example,
\ba
\tilde{\Psi}_{+} \Psi_{+} & = & (a_{+}a_{+}^{*} + b_{+}b_{+}^{*} )E_{+ 3} \ , \nonumber \\
\tilde{\Psi}_{+} {\bf e}_{3} \Psi_{+} & = & (a_{+}a_{+}^{*} - b_{+}b_{+}^{*} )E_{+ 3} \ . \nonumber \\
\ea
So upon interpretation of these ideals as spinor-wavefunctions we can associate to the coefficients $a_{+}$ and $b_{+}$ the intepretation as probabillity amplitudes for spin-up and spin-down states. This algebra suggests that {\it non-relativistic} spin $\frac{1}{2}$ fermions come in doublets as we have two such ideals, $I_{\pm}$. Yet the fact that the projector $E_{+3}$ is part of the expectation values is ugly and indicates that we are missing some crucial element. In the following section we will see that once we set up {\it relativistic} Pauli ideals, i.e. ideals $I_{\pm}$ in the even-grade subalgebra of STA this problem is solved.

\newsection{Pauli Ideals in Spacetime algebra}
We will work in a Minkowski spacetime with signature $-2$. In STA the generators of Lorentz boosts satisfy a Pauli algebra just like the vectors in 3d GA. Together with the generators of rotations they satisfy a complexified Pauli algebra, just like vector + bivectors do in 3d, however in spacetime all generators are of the same grade. Although the spacetime peudoscalar $i$ anti-commutes with vectors, it commutes with all bivectors and thus allows us to go through the same steps as in 3d. Due to the fact that single-sided Lorentz transformations preserve the the Pauli-ideals from the even-grade subalgebra, the even-grade spinors fall apart into Lorentz invariant classes. This suggests that these could be physically distinguishable. In this section we will construct these Pauli-ideals and compute the corresponding "expectation values".

\newsubsection{Generating the ideals $I_{\pm}$ }
Let the generators of the even-grade sub-algebra of STA be the timelike bivectors $K_{j} = {\bf e}_{j} {\bf e}_{0}$, $j=1,2,3$. We find that the corresponding ideals $I_{\pm}$ are generated by
\ba
E_{\pm 3} & = & \frac{1}{2} ( 1 \pm K_{3} ) \ , \\
D_{\pm 3} & = & K_{1}E_{\pm 3} = \frac{1}{2} ( K_{1} \mp i K_{2} ) \ ,
\ea
where $i$ now refers to the spacetime pseudoscalar. Note that $E_{\pm 3} $ is a projector. The multiplication of any of these two with a $K_{j}$, or with a spacelike bivector $J_{k}$, from the left, returns one of the generators again, possibly multiplied by the pseudoscalar. As a result, the spinors
\be
\Psi_{\pm} = a_{\pm} E_{\pm 3} + b_{\pm} D_{\pm 3} , 
\ee
can again represent spinor-wavefunctions in the same way as before. From the above definitions it is easy to check that for a given even-grade multivector $\Psi$ its decomposition into representatives from the ideals $\Psi_{\pm}$ satisfies
\be
\Psi_{\pm} = \Psi E_{\pm 3} \ ,
\ee
and
\ba
E_{\pm 3} \Psi_{\pm} & = & a_{\pm} E_{\pm 3} \ , \nonumber \\
E_{\mp 3} \Psi_{\pm} & = & b_{\pm} D_{\pm 3} \ . \nonumber
\ea
The algebra is considerably simplified by a set of identities satisfied by these generators, such as
\ba
E_{\pm 3}^2 = E_{\pm 3} , & D_{\pm 3}^2 = 0 , & E_{\mp 3} D_{\pm 3} = D_{\pm 3} \ , \nonumber \\
E_{\pm 3} D_{\pm 3} = 0 , & D_{\mp 3} E_{\pm 3} = 0 , & D_{\pm 3} E_{\pm 3} = D_{\pm 3} \ , \\
E_{\pm 3} E_{\mp 3} = 0 , & {\widetilde{D_{\pm 3}}} = -D_{\mp 3}  , & {\widetilde{E_{\pm 3}}} = E_{\mp 3}  \ . \nonumber
\ea
Their validity only depends on the manner in which the ideals are constructed, but not on the choice of representation. All algebra below only requires the above relations. Finally, also of interest are
\ba
{\bf e}_{0} E_{\pm 3}  & = & E_{\mp 3}  {\bf e}_{0} \ , \nonumber \\
{\bf e}_{0} D_{\pm 3}  & = & - D_{\mp 3}  {\bf e}_{0} \ . \nonumber 
\ea
The parity transformation in STA is ofcourse defined with respect to a given frame. In the present frame an arbitrary multivector $M$ transforms under parity as
\be
M' = {\bf e}_{0} M {\bf e}_{0} \ .
\ee
If we compare this to the previous equation we see that parity swaps the ideals, but also that the $E$ and $D$ components of an ideal transform {\it distinctly} under parity. The Pauli-ideals are invariant under left-multiplying Lorentz transformations. Their physical significance follows from the view that physical states of an elementary physical process should form an irreducible representation of the Lorentz transformations. 

Due to the overall even-grade of the $\Psi_{\pm}$ we can use them to construct grade-preserving linear maps on multivectors $M$ by bilinear sandwiching $\Psi_{\pm} M \tilde{\Psi}_{\pm}$. Let the weighed rotor $\Psi$ represent the state of a particle, and some multivector $M$ a physical/geometric property of this particle in it's particle-frame. This frame cannot in general be a rest frame as, as we will see, we are contemplating massless modes.  Its value in the observers' frame ${\bf e}_{\mu}$ is then given by
\be
\langle M \rangle = \Psi M {\tilde{\Psi}}  \ ,
\ee
called the {\ expectationvalue} of $M$. This expression holds both classically as well as quantum mechanically. Care should be taken when applying this to the seperate ideals as
\be
\Psi_{\pm} {\tilde{\Psi}}_{\pm} = 0 \ , 
\ee
and so they are not regular {\it weighed} rotors as we know them. In particular we'll have
\be
\langle A B \rangle \neq \langle A \rangle  \langle B \rangle \ .
\ee

\newsubsection{Expectation values and flavour $SU(2)$}
If we write
\ba
a_{\pm} & = & r_{\pm a} \exp{(i \theta_{\pm a})} \ , \nonumber \\
b_{\pm} & = & r_{\pm b} \exp{(i \theta_{\pm b})} \ , \nonumber 
\ea
we can compute the expectation values
\ba
{\bf j}_{\pm} & \equiv & \Psi_{\pm} {\bf e}_{0} \tilde{\Psi}_{\pm} \ , \nonumber \\
& = & \frac{1}{2} \left(r_{\pm a}^2+r_{\pm b}^2\right) {\bf e}_{0} + r_{\pm a} r_{\pm b} \cos (\theta_{\pm a}-\theta_{\pm b}) {\bf e}_{1} \nonumber \\
& & \mp r_{\pm a} r_{\pm b} \sin (\theta_{\pm a} - \theta_{\pm b}) {\bf e}_{2}   \pm \frac{1}{2} \left(r_{\pm a}^2-r_{\pm b}^2\right) {\bf e}_{3} \ .
\ea
The ${\bf j}_{\pm}$ is a current-density. The projectors $E_{\pm 3}$ are no longer part of the expectation values. The vectorfield ${\bf j}_{\pm}$ is unique for $\Psi_{\pm}$ in the sense that the expectation values of ${\bf e}_{1}$ and ${\bf e}_{2}$ vanish and the expectation value for ${\bf e}_{3}$ equals 
\ba
{\bf s}_{\pm 3} & \equiv & \Psi_{\pm} {\bf e}_{3} \tilde{\Psi}_{\pm} \ , \nonumber \\
& = & \pm {\bf j}_{\pm} \ .
\ea
Hence there is a current ${\bf j}_{\pm}$ for each ideal and the second current ${\bf s}_{\pm 3}$ is either parallel {\it or} anti-parallel. As it would be natural to associate ${\bf j}_{\pm}$ with the particle-number current-density of the corresponding field $\Psi_{\pm}$, the behaviour of the the current-density ${\bf s}_{\pm 3}$ seems to indicate the different ideals carry some charge of opposite sign. The analysis of the other expectation values coroborates this idea.

In general a weighed rotor will be a superposition of elements from both ideals, i.e. $\Psi = \Psi_{+} + \Psi_{-}$. If we compute the expectation values for such a rotor we find
\ba
\Psi {\bf e}_{0} \tilde{\Psi} & = & \Psi_{+} {\bf e}_{0} \tilde{\Psi}_{+} + \Psi_{-} {\bf e}_{0} \tilde{\Psi}_{-} = {\bf j}_{+} + {\bf j}_{-}  \ , \nonumber \\
\Psi {\bf e}_{3} \tilde{\Psi} & = & \Psi_{+} {\bf e}_{3} \tilde{\Psi}_{+} + \Psi_{-} {\bf e}_{3} \tilde{\Psi}_{-} = {\bf j}_{+} - {\bf j}_{-}   \ , \\
\Psi {\bf e}_{j} \tilde{\Psi} & = & \Psi_{+} {\bf e}_{j} \tilde{\Psi}_{-} + \Psi_{-} {\bf e}_{j} \tilde{\Psi}_{+}  \ , \ j = 1,2 \ . \nonumber 
\ea
Obviously $\Psi {\bf e}_{0} \tilde{\Psi}$ no longer needs to be a null-vector. The above results are most easilly interpreted as a phenomenon originating from a flavor $SU(2)$ symmetry. This works as follows, first we note that
\ba
{\bf e}^{j} = {\bf e}^{0} K_{j} \ , \nonumber
\ea
so that we can write
\be
{\bf \s}^{\pm j} \equiv \Psi_{\pm} {\bf e}_{0} K_{j} \tilde{\Psi}_{\pm} \ .
\ee
Now we interpret the $K_{j}$ as {\it charge-operators} $2 Q_{j}$, the three operators satisfying the $SU(2)$ commutation relations by construction. Now the expressions derived above can be regarded as implying that the $Q_{3}$-charge of the rotors $\Psi_{\pm}$ satisfies  $Q_{3} = \pm \frac{1}{2}$. Pure ideals satisfy $Q_{1,2} = 0$, but in general a weighed rotor is a superposition of such states. An active rotation in the particle's frame acts on the weighed rotor representing the particle's state as
\be
\Psi' = \Psi e^{-\frac{i}{2}\a_{j} K_{j}} = \Psi e^{-i \a_{j} Q_{j}} \ .
\ee
This is suggestive of a weak-interaction interpretation of these expectation values. However, formalism hardly ever restricts itself to a single interpretation, and so in the final subsection we also review the expectation values for bivector quantities which could be though of as dipole moments of some kind.

\newsubsection{Dipole moments}
The bivector expectation value is similarly unique modulo dualisation and reads
\be
K_{\pm} \equiv \Psi_{\pm} K_{1} \tilde{\Psi}_{\pm} \ . \nonumber
\ee
As the $\Psi_{\pm}$ are of even-grade they commute with the unit-pseudoscalar and consequently we have
\be
\Psi_{\pm} K_{j} \tilde{\Psi}_{\pm} = - i \Psi_{\pm} J_{j} \tilde{\Psi}_{\pm} \ .
\ee
So the 6 possible values reduce to 3, but due to the fact that ${\bf e}_{0}$ and ${\bf e}_{3}$ generate parellel, or anti-parallel, lightlike expectation values it will not be surprising that 
\be
\Psi_{\pm} K_{3} \tilde{\Psi}_{\pm} = 0 \ .
\ee
This leaves the expectation values for $K_{1}$ and $K_{2}$. Straightforward computation reveals that
\be
\Psi_{\pm} K_{1} \tilde{\Psi}_{\pm} = \pm i \Psi_{\pm} K_{2} \tilde{\Psi}_{\pm} \ ,
\ee
implying there is for each ideal one expectation value, modulo dualisation, on the bivecor level. I find for the $\Psi_{+}$ case
\ba
K_{+} & = & \frac{1}{2} \{ r_{a}^2 \cos{(2 \th_{a})} - r_{b}^2 \cos{(2 \th_{b})} \} K_{1} - \frac{1}{2} \{ r_{a}^2 \cos{(2\th_{a})}  + r_{b}^2 \cos{(2 \th_{b})} \} K_{2} \nonumber \\ 
&  &  - r_{a}r_{b} \cos{(\th_{a} + \th_{b})} K_{3} +  r_{a}r_{b} \cos{(\th_{a} + \th_{b})} J_{3} \\
& & - \frac{1}{2} \{ r_{a}^2 \cos{(2 \th_{a})} - r_{b}^2 \cos{(2 \th_{b})} \} J_{1} - \frac{1}{2} \{ r_{a}^2 \cos{(2\th_{a})}  + r_{b}^2 \cos{(2 \th_{b})} \} J_{2} \ . \nonumber 
\ea
The {\it bivector}-expectationvalue is a null-bivector. When we take it to describe the electromagnetic dipole-moment of a massless particle, the four distinct cases ,$\{K_{+} ,i K_{+} ,K_{-} , i K_{-} \}$, allow a choice between, or a mixture of, electric- and magnetic-dipole moments. 

For particles whose description requires a superposition of both ideals we find similar behaviour to what we found earlier for the currents. The vanishing expectation values of $J_{3}$ and $K_{3}$ are now finite due to the mixing of ideals whereas the contributions from each ideal to the expectation values of $K_{1,2}$ are simply additive.

\newsection{Dirac Fermions}
We can apply the above notions to what we known about the algebra of Dirac matrices. The massless Dirac equation can be describes two physically distinct excitations: a left-handed and a right-handed fermion. These modes enter into the electroweak interactions in a distinct manner also including a second fermion; the neutrino. Naively one could thus expect a need for {\it two} copies the Pauli-ideals and not just one. The idea of having two particles described by a single mathematical object in this context requires us to contemplate the idea of having two rotors instead of one. In such a construction the algebra of the Dirac matrices should be represented using $2 \times 2$-matrices with STA insertions. Such a construction is straightforward.

\newsubsection{Bi-rotors}
We are going to consider matrices of rotors, i.e.
\be
| {\bf \Psi} > = \twomatrix{\psi_{1}}{\phi_{1}}{\psi_{2}}{\phi_{2}} \ ,
\ee
in which all entries $\psi_{1}, \psi_{2}, \phi_{1}, \phi_{2}$ are weighed rotors, i.e. even-grade multivectors from STA. Now define the $\g$-matrices in the chiral representation as,
\ba
\g_{0} & = & \twomatrix{0}{1}{1}{0} \ , \\
\g_{j} & = & \twomatrix{0}{K_{j}}{-K_{j}}{0} \ .
\ea
This representation is chiral because the rotor-fields $\psi_{1}$ and $\psi_{2}$ will turn out to form a parity-pair. We want to associate the matrix of weighed rotors $|{\bf \Psi}>$  with particular linear combinations of the $\g$-matrices with coefficients of scalar + pseudoscalar nature. A useful set of basis elements are the generators
\ba
E_{DR3}^{\pm} & = & \frac{1}{4} \{ [ \g_{0}^2 + i \g_{0} \g_{1} \g_{2} \g_{3} ] \pm [ -\g_{3}\g_{0}  + i \g_{1} \g_{2} ] \} \ , \nonumber \\
E_{UL3}^{\pm} & = & \frac{1}{4} \{ [ \g_{0}^2 - i \g_{0} \g_{1} \g_{2} \g_{3} ] \pm [ \g_{3}\g_{0}  + i \g_{1} \g_{2} ] \} \ , \nonumber \\
E_{UR3}^{\pm} & = & \frac{1}{4} \{ [ \g_{0} + i \g_{1} \g_{2} \g_{3} ] \pm [ \g_{3}  + i \g_{0} \g_{1} \g_{2} ] \} \ , \nonumber \\
E_{DL3}^{\pm} & = & \frac{1}{4} \{ [ \g_{0} - i \g_{1} \g_{2} \g_{3} ] \pm [ -\g_{3}  + i \g_{0} \g_{1} \g_{2} ] \} \ .  
\ea
They put a generator $E_{3\pm}$ in each of the four entries of the two-by-two matrices. The extra 'indices' have prozaic meanings like $UL = upper-left$ indicating the corresponding entry in the matrix. Ofcourse more speculative, and possibly appealing, names such as "upper-lepton" or "lower-quark" are also possible. Similarly the set
\ba
D_{DR3}^{\pm} & = & \frac{1}{4} \{ - [\g_{1} \g_{0} - i \g_{2} \g_{3} ] \pm [ \g_{3}\g_{1}  - i \g_{2} \g_{0} ] \} \ , \nonumber \\
D_{UL3}^{\pm} & = & \frac{1}{4} \{ [\g_{1} \g_{0} + i \g_{2} \g_{3} ] \pm [ \g_{3}\g_{1}  + i \g_{2} \g_{0} ] \} \ , \nonumber \\
D_{UR3}^{\pm} & = & \frac{1}{4} \{ [ \g_{1} + i \g_{0} \g_{2} \g_{3}] \pm [ \g_{0} \g_{3} \g_{1}  - i \g_{2}] \} \ , \nonumber \\
D_{DL3}^{\pm} & = & \frac{1}{4} \{-[ \g_{1} - i \g_{0} \g_{2} \g_{3}] \pm [ \g_{0} \g_{3} \g_{1}  + i \g_{2}] \} \ . 
\ea
All together sixteen generators that can be linearly combined using scalar-pseudoscalar coefficients. This set evidently generates the right number of degrees of freedom. 

We can clearly see what happens if, at this point, we identify the Dirac $\g$-matrices with orthonormal basis-vectors of STA. Precisely half of the generators just defined vanishes, and as a result half of the ideals does. The remaining generators split into odd-grade and even-grade multivectors, which would then correspond to the even- and odd-grade parts of the resulting spacetime algebra. For example consider applying the linear map defined by
\be
\g_{\mu} \rightarrow {\bf e}_{\mu} \ .
\ee
It leads to
\ba
E_{DR3}^{\pm} = 0 & , &  D_{DR3}^{\pm} = 0 \ ,\nonumber \\
E_{UL3}^{\pm} = E_{\pm 3} & , &  D_{UL3}^{\pm} = D_{\pm 3} \ , \nonumber \\
E_{UR3}^{\pm} = E_{\pm 3} {\bf e}_{0} & , & D_{UR3}^{\pm} = D_{\pm 3} {\bf e}_{0}  \ , \nonumber \\
E_{DL3}^{\pm} = 0 & , & D_{DL3}^{\pm} = 0  \ . \nonumber  
\ea
As a result the two particles described by the weighed rotors $\psi_{1}$ and $\phi_{1}$ are now combined into a single multivector that, with a slight abuse of notation, we can write as
\be
| {\bf \Psi} > \ \rightarrow \psi_{1} + \phi_{1} {\bf e}_{0} \ .
\ee
Other mappings $\g \rightarrow {\bf e}$ can also be chosen to isolate different pairs of rotor-fields,
\ba
\g_{\mu} \rightarrow {\bf e}^{\mu}:& & \ \ | {\bf \Psi} > \ \rightarrow \psi_{2} + \phi_{2} {\bf e}_{0} \ , \\
\g_{0,3} \rightarrow {\bf e}^{0,3} \ , \ \g_{1,2} \rightarrow {\bf e}^{2,1} :& & \ \ | {\bf \Psi} > \ \rightarrow \psi_{1} + \psi_{2} {\bf e}_{0} \ .
\ea
The last of the two shows the parity-pair from standard Dirac theory to appear as a full (i.e. both even and odd-graded) spinorial multivector. This is the pairing also found by Hestenes in \cite{sm}.

The above examples are just particular cases of the possibillity of using a general linear map 
\be
\g_{\mu} = L_{\mu}^{\nu} {\bf e}_{\nu} \ .
\ee
Note that the chiral projection operator 
\be
P_{\pm C} = \frac{1}{2} \{1 \pm i \g_{5} \} \ ,
\ee
under orthogonal maps between the orthonormal bases $\g_{\mu}$ and ${\bf e}_{\nu}$ decouples the chiral ideals. There is ofcourse nothing against having a second copy of STA where the alternate choice has been made. We could allow the coefficients $L_{\mu}^{\nu}$ to be scalar-peudoscalar combinations. As a result there would be a "hidden" $U(4)$-symmetry the the mapping between STA and the Dirac algebra. 

We have constructed a Dirac algebra in the chiral representation from $2 \times 2$-matrices with STA entries in a chiral representation. We could have used an entirely different representation and draw the same conclusions. The representation of the Dirac algebra as $2 \times 2$-matrices with STA entries suggests it results from having {\it two} basis-frames instead of just one. As a final step in this section let us take a look at the Dirac equation and see how the four spinors fit into that picture.

\newsubsection{The Dirac Equation}
Using that $\g_{0} = \g^{0}$ and $\g_{j} = - \g^{j}$ The Dirac operator reads
\be
\nabla_{D} = \sum_{\mu = 0}^{3} \g^{\mu} \del_{\mu} \ .
\ee
This can be rewritten using the ordinary vectorderivative $\nabla$
\be
\nabla = \sum_{\mu = 0}^{3} {\bf e}^{\mu} \del_{\mu} \ ,
\ee
as
\be
\nabla_{D} = \twomatrix{0}{ {\bf e}_{0} \nabla }{ \nabla {\bf e}_{0}}{0} \ .
\ee
The STA-Dirac equation reads
\be
\nabla_{D} | {\bf \Psi} > + m | {\bf \Psi} > \hat{S}_{0} = 0 \ ,
\ee
where $\hat{S}_{0}$ is the spacelike unit-bivector representing the {\it spinplane} of the particle. This set of two equations can also be derived classically when making the transition from an eigen-rotor description of particles to a description in terms of a rotor-field \cite{clasdirac}. Infact, there the whole origin of the ideals can be attributed to the appearance of $\hat{S}_{0}$ as we could have chosen $K_{3} = -i \hat{S}_{0}$.

As the Dirac operator operates from the left, the collumns in the $2 \times 2$-matrix state $| {\bf \Psi} > $ are preserved. This Dirac equation is the wellknown two-component form of the Dirac equation, but there is a substantial difference! The spinors $\psi_{j}$ are each  doublets of 2-spinors. The Dirac equation will not mix the flavour doublets if for the "generator" of the ideals we choose $K_{3} = -i \hat{S}_{0}$. The standard Dirac theory parity transformation
\be
| {\bf \Psi} >^{P} =  \g_{0} | {\bf \Psi} > ,
\ee
swaps the two fields $\psi_{1}$ and $\psi_{2}$ identifying them as the left- and righthanded fermion component. In the weak interactions the two components enter differently.

\newsection{Summary and Discussion}
It is possible to represent the Dirac $\g$-matrices using spacetime algebra in the form of $2 \times 2$-matrices with bivector-entries.  Dirac spinors can then be represented by a collumn of a $2 \times 2$-matrix with weighed-rotors as insertions. This representation has the correct number of ideals while remaining {\it real}. Each of the weighed rotors {\it can} be chosen to represent an $SU(2)$-flavour doublet, and the action of the corresponding flavour rotations is merely a rotation in the particle-frame. 

The main difference of the work presented here to that presented elswhere, in particular in \cite{sm}, is that here we represent the two 2-spinors of a massless left-handed fermionic $SU(2)$ doublet in a single rotor-field. In contrast \cite{sm} puts the two chiral 2-spinors  of a fermion field into a single rotor-field. We have show that this difference reflects, in part, a difference in choice of the mapping $\g_{\mu} \rightarrow {\bf e}_{\nu}$. However one could argue that the two chiral components are, at least on the level of weak interactions, basically very distinct degrees of freedom.

Let $\psi_{1}$ and $\psi_{2}$ be two such rotors in a collumn. If we make the identifications for electron and neutrino according to
\ba
e_{L} = \psi_{-1} \ , \nu_{L} = \psi_{+1} \ , \nonumber \\
\Psi_{L} = e_{L} + \nu_{L} \ ,
\ea
this requires
\ba
e_{R} = \psi_{-2}  \ , \psi_{+2} = 0 \ , \nonumber \\
\Psi_{R} = e_{R} \ ,
\ea
for the standard model with massless neutrino's. Under electroweak $SU(2)$ the rotor-fields transform as under "restframe"-rotations. In the minimal SM the Higgs-field $\Phi$ is a Lorentz-scalar and a iso-spinor i.e. it transforms according to
\ba
\Psi_{L}' & = & \Psi e^{-\frac{1}{2}\a_{k} J_{k}}  \ , \nonumber \\
\Phi' & = &  e^{\frac{1}{2}\a_{k} J_{k}} \Phi \ . 
\ea
This set up adorns the bivector $i \hat{S}_{0}$ with an association {\it not} to the electromagnetic charge \cite{sm}, but rather to a diagonalised fermion flavour charge. To see whether all this gives any insight into the origin of the Higgs field we need to look at a generic Higgs-Fermion coupling.

The standard $SU(2)$-symmetric coupling is
\be
iG_{H} \{ \Phi^{\dagger} \Psi_{L}^{\dagger} \Psi_{R} + \Psi_{R}^{\dagger} \Psi_{L}\Phi   \} \ . 
\ee
Although such a term in the Langrangian is Lorentz-scalar, it is not algebraically scalar. An analysis of the classical roots of the Dirac equation reveals bivector-valued actions to be unproblematic, even classically. These results will be reported elsewhere \cite{clasdirac}.
The Higgs field also is some kind of doublet $\Phi = \{\phi_{1},\phi_{2}\}$ of scalar-pseudoscalars, and the spontaneous symmetry breaking that occurs in the Higgs sector must yield
\ba
\Psi_{L} \Phi = e_{L}\phi_{1} + \nu_{L} \phi_{2} = - \varphi_{0} e_{L} K_{3} \ ,
\ea
where $\varphi_{0}$ is the scalar vacuum expectation value of the Higgs field. Due to the fact that the $\Psi_{L}$ were constructed as left-ideals by projecting from the right with $\frac{1}{2}(1 \pm K_{3})$ we automatically have
\ba
e_{L} = e_{L} K_{3} \ .
\ea
But instead of associating the Higgsfield with the fairly meaningless and arbitrary $K_{3}$ the above suggests the condensated Higgs field is infact {\it the projector} $E_{\pm3}$ themselves! The four real degrees of freedom of the Higgs field can be assembled as
\be
\Phi = \phi_{1} E_{-3} + \phi_{2} E_{+3} = (\phi_{1} + \phi_{2}) + (\phi_{1} - \phi_{2}) K_{3} = \varphi ( 1 + \chi K_{3} ) \ ,
\ee 
where the Lorentz scalars $\varphi$, and $\chi$ are the obvious linear combinations of the $\phi_{j}$. The $\b$ can be gauged away as a result of another $U(1)$ gauge freedom related to hypercharge. The origin of this symmetry can also be traced in the classical form of the Dirac equations \cite{clasdirac}. If the lagrangian of the Standard Model contains the ordinary symmetry-breaking potential 
\ba
V_{\Phi} & = & \frac{m^2}{2} \Phi^{\dagger} \Phi + \frac{\l}{4} (\Phi^{\dagger} \Phi )^{2} \ , \nonumber 
\ea
the resulting expression will be Lorentz-scalar and hermitian but algebraically contain the bivector $K_{3}$. This makes it worthwhile to contemplate whether an alternative symmetry-breaking lagrangian exists that makes explicit use of the possible geometric character of the Higgsfield and would give a geometric requirement for $\Phi$ to break down into a projector. Work on these matters is in progress.

\end{document}